# HOW LOW - ENERGY FUSION CAN OCCUR

B. Ivlev

**Fusion of two deuterons of room temperature energy is discussed. The nuclei are in vacuum with no connection to any external source (electric or magnetic field, illumination, surrounding matter, traps, etc.) which may accelerate them. The energy of two nuclei is conserved and remains small during the motion through the Coulomb barrier. The penetration through this barrier, which is the main obstacle for low-energy fusion, strongly depends on a form of the incident flux on the Coulomb center at large distances from it. In contrast to the usual scattering, the incident wave is not a single plane wave but the certain superposition of plane waves of the same energy and various directions, for example, a convergent conical wave. The wave function close to the Coulomb center is determined by a cusp caustic which is probed by de Broglie waves. The particle flux gets away from the cusp and moves to the Coulomb center providing a not small probability of fusion (cusp driven tunneling). Getting away from a caustic cusp also occurs in optics and acoustics.**

1. GENERAL VIEW

Fusion of atomic nuclei may release a lot of energy. The aspects of nuclear fusion are discussed, for instance, in Refs. [1-10] and references therein. To get at a short nuclear distance, where fusion occurs, nuclei should overcome a high Coulomb barrier. This is the main obstacle to get fused. There are two ways to pass the Coulomb barrier, either to accelerate the nuclei up to high energy, comparable with the barrier height (of the order of 1MeV), or to pass the barrier via quantum tunneling. When the energy is not high, the probability of tunneling of the nuclei through the Coulomb barrier is extremely small according to the theory of Wentzel, Kramers, and Brillouin (WKB) [11]. So only high energy nuclei can fuse.

As an example one can consider the fusion reaction of two deuterons

$$^{2}_{1}H + ^{2}_{1}H \rightarrow ^{3}_{2}He + n + 3.27\,MeV \qquad (1)$$

which relates to the tunneling process shown in Fig. 1. The probability of barrier penetration depends on the deuteron mass $M$ and is proportional to $\exp(-2\pi\lambda/r_B)$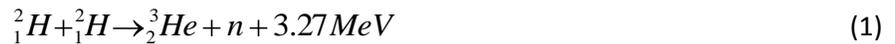, where $\lambda = \hbar/\sqrt{ME}$ specifies the de Broglie wave length of the deuteron and $r_B = 2\hbar^2/Me^2 \approx 2.9 \times 10^{-12}\,cm$ is the nuclear Bohr radius. When the energy of two deuterons is $E \rightarrow 300\,Kelvin$ the tunneling probability is proportional to $10^{-2682}$ in accordance with usual estimates for low-energy fusion. The Coulomb scale is $e^2/E \approx 500\,A°$. To get the probability of a reasonable value one has to use high energy nuclei with the energy of the order of $10^8\,Kelvin$ which corresponds to the barrier height in Fig. 1.

Therefore, high energy nuclei are the leading idea of fusion technique. Toroidal design of high-energy plasma (TOKAMAK) is a way to get controlled nuclear fusion. See, for example, [1]. Another direction is the use of cold systems which are only locally "hot". An example is an acceleration of initially cold deuterons by a strong electric field using a pyroelectric crystal [5].

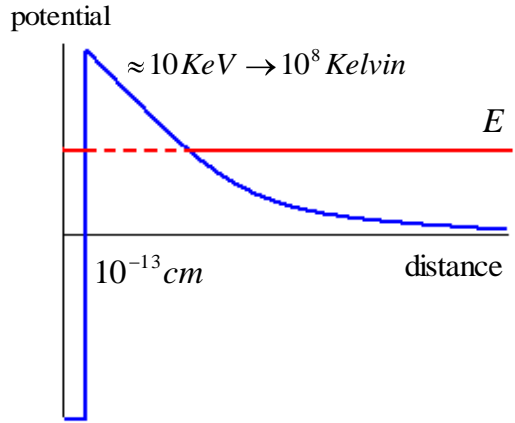

**Figure 1.** Penetration trough the Coulomb barrier according to the usual tunneling mechanism. Here $E$ is a particle energy. The Coulomb potential is extended down to the nuclear size.

It is surprising to claim that *two bare deuterons*, that is isolated from everything, *of room temperature energy* are able, in principle, to penetrate the Coulomb barrier with a not small probability and to subsequently fuse. The statement is counterintuitive. A tennis ball cannot pass through a brick wall. This correlates with the usual underbarrier physics led by the philosophy of addition of probabilities but not amplitudes. The proposed phenomenon of barrier penetration is based ultimately on interference that is on addition of amplitudes.

The paper deals with two low-energy nuclei in vacuum. There are no external sources (electric or magnetic field, illumination, surrounding matter, traps, etc.) which may accelerate them. The energy of the two nuclei, in the system of center of mass, is conserved and remains small during the motion through the Coulomb barrier. This is a substantial difference from usual schemes to push in action a mechanism of low-energy fusion by some local heating or acceleration of nuclei.

The presented paper is a short version of detailed Ref. [12]. See also related publications [13 -15]. We are restricted here by general arguments supported by figures.

2. CLASSICAL MECHANICS

In this section we consider classical trajectories of particles approaching the Coulomb center. A situation strongly depends on a form of the incident flux.

(a) Parallel flux on the Coulomb center

The usual formulation of a scattering problem is connected with an incident plane wave on the Coulomb center. There is a classical analogue of the quantum mechanical process when parallel classical trajectories are reflected by the Coulomb field. This is illustrated in Fig. 2(a). The further behavior of scattered trajectories is shown in Fig. 2(b) when they are reflected from the certain surface named caustic [16].

(b) Conical flux on the Coulomb center.

In the absence of the Coulomb center the classical trajectories of the conical flux are shown in Fig. 3(a). This flux distribution is not exotic since it can be realized, for example, in a cylinder demonstrated in Fig. 3(b). The conical flux is supposed to be axially symmetric.

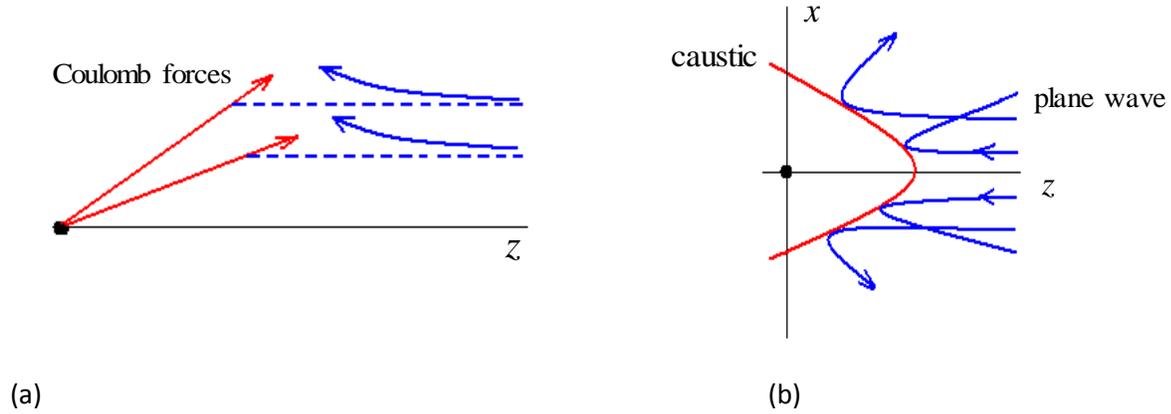

(a)                        (b)

**Figure 2**. (a) Deviation of classical parallel trajectories by the Coulomb forces. (b) Classical trajectories are reflected from the axially symmetric surface named caustic. The intersection of this surface and the plane $\{x, z\}$ is shown.

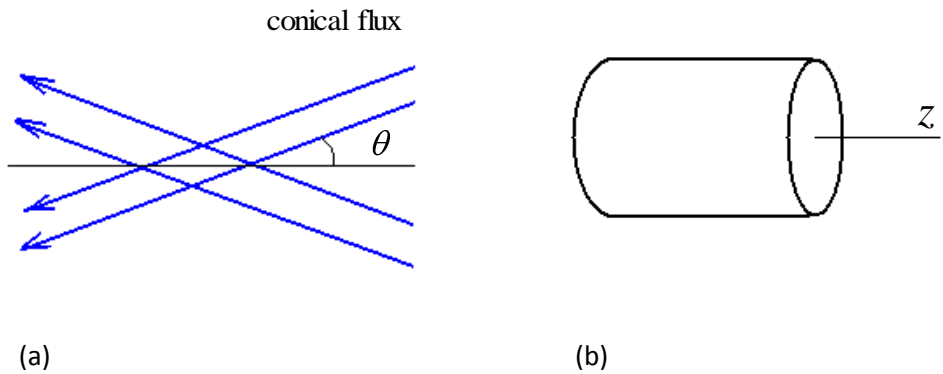

(a)                        (b)

**Figure 3**. (a) Classical trajectories related to the conical flux in the absence of the Coulomb center. (b) Particular realization of the conical flux in a cylinder with reflection from its surface.

If we add the Coulomb center the conical distribution in Fig. 3(a) will be substantially deformed. It is clear from Fig. 4 where the trajectories apart from the direction of the angle $\theta$ are deviated to opposite sides. Analogously to parallel flux, the deviations give rise to the caustics $A$ and $B$ shown in Fig. 5. These caustics are continued from the infinity toward the Coulomb center which remains in the shadow of those caustics. The caustic $A$ is of a cusp form as occurs for convergent incident flux. The cusp position is $z_s = e^2 / E \sin^2 \theta$ [12]. The analogous situation with light rays is shown in Fig. 6(a) where rays become convergent after reflection from the circle mirror.

The incident trajectory 1 in Fig. 5 is reflected from the caustic $A$, intersects the $z$ axis, and goes over into the trajectory 2 which pierces the caustic $A$. The rays of light, analogous to trajectories 2, are shown in Fig. 6(a) to the left from the caustic. Unlike straightforward light rays, trajectories 2 behave differently due to a deviation in the Coulomb field. As a result, they are reflected from the certain caustic $C$ in Fig. 5.

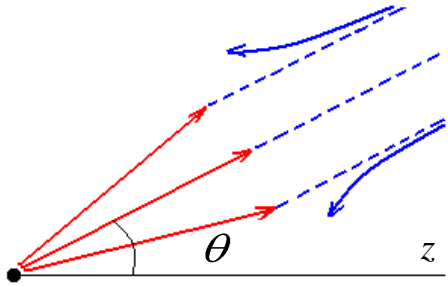
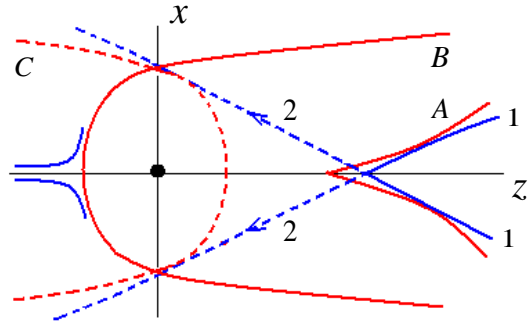

**Figure 4.** Deviation of classical trajectories from the conical distribution by the action of Coulomb forces.

**Figure 5**. Reflection of the incident conical flux from caustics. The cusp position is $z_s$.

Trajectories in Figs. 2-5, and also 6(a) correspond to classical physics. The Coulomb center is in the absolute shadow of all caustics. Otherwise it would be strange to get the Coulomb center reachable within classical mechanics.

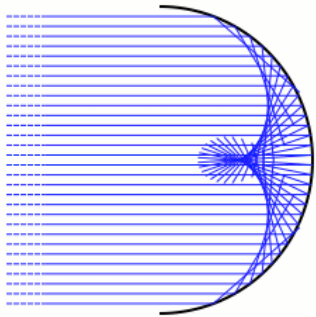
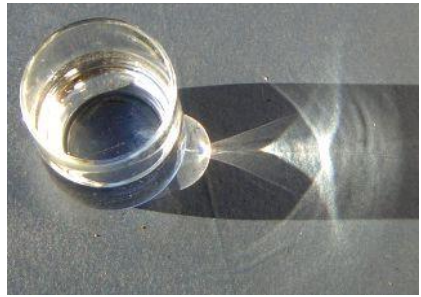

(a)                                (b)

**Figure 6**. (a) Geometrical optics. Parallel rays are reflected from the circle, go over into the convergent flux, and leave the caustic in the classical manner just piercing it. (b) Wave optics. Formation of the cusp caustic in reality.

3. WAVE FUNCTION

In formation of cusp caustics geometrical optics, Fig. 6(a), goes over into wave optics, Fig. 6(b), when the wave length is finite. This is similar to a transition between classical and quantum mechanics [11].

All classical trajectories in Fig. 5 go apart from the Coulomb center of course. How quantum mechanics can violate this scenario? To the left from the caustic $A$ there are two contributions of different nature, $\psi_2$ and $\psi_1$, to the total wave function. This is marked in Fig. 7.

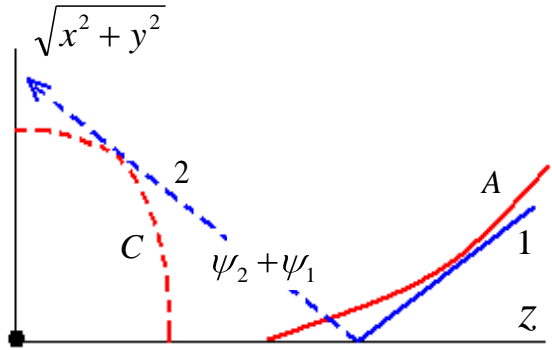

**Figure 7**. Formation of wave functions $\psi_2$ (rays state) and $\psi_1$ (cusp driven state) to the left from the caustic $A$ is described in the text. The Coulomb center is in the absolute classical shadow.

(a) Rays state $\psi_2$

According to Feynman, a wave function in quantum mechanics can be considered as a continuous superposition of classical trajectories. In our case the wave function $\psi_2$ relates to the set of classical trajectories (rays, analogously to optics) 2 in Fig. 8(a). The rays state $\psi_2$ is similar to one corresponding to classical trajectories in Fig. 2(b). In the both states the momentum, perpendicular to the $z$ axis, is zero on this axis. Therefore these states are similar to one-dimensional ones when the momentum, parallel to the $z$ axis, is zero at the classical turning point. That is the caustic in Fig. 2(b) and the caustic $C$ in Fig. 7 intersect the $z$ axis when the condition $e^2/z = E$ holds. The state $\psi_2$ exponentially decays to the left from the caustic $C$. The classical analogue of the rays state $\psi_2$ corresponds to Fig. 6(a).

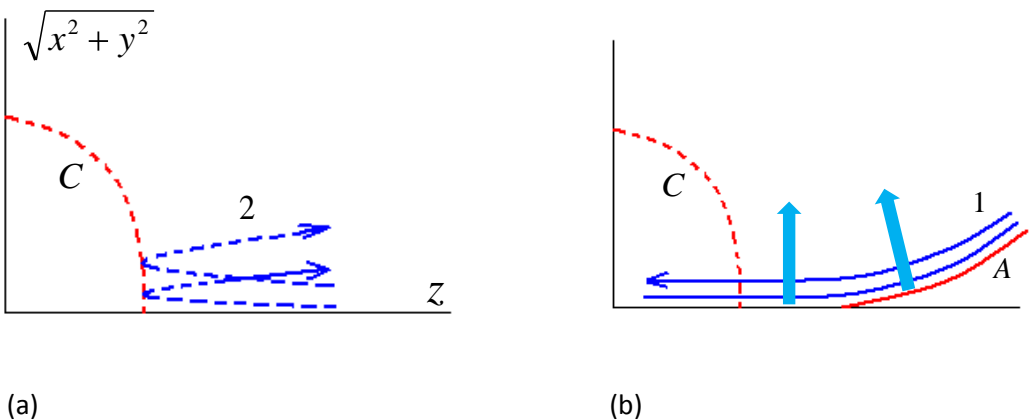

(a)            (b)

**Figure 8**. (a) Rays state $\psi_2$ relates to the set of classical trajectories 2. $\psi_2$ exponentially decays to the left from the caustic $C$. (b) The wave function $\psi_1$ of the cusp driven state exponentially decays inside the shadow region (along the thick arrows) of the caustic $A$.

(b) Cusp driven state $\psi_1$

In contrast to classical physics, a caustic shadow is not completely "empty" due to an exponential decay of the wave function inside the shadow region [11, 16]. Accordingly, the certain state $\psi_1$ exponentially decays along the thick arrows in Fig. 8(b). Not far from the cusp point features of this cusp driven state are given by the formula [12]

$$\psi_1 \sim \exp\left(-\frac{iz}{\lambda}\cos\theta - \sin\theta \frac{\sqrt{x^2+y^2}}{\lambda}\sqrt{1-\frac{z}{z_s}}\right). \quad (2)$$

At $z = 0$ and not far from the Coulomb center

$$|\psi_1| \sim \exp\left[-\frac{2(x^2+y^2)^{1/4}}{\sqrt{r_B}}\right]. \quad (3)$$

In equations (2) and (3) arguments should be formally large. They become of the order of unity at the border of applicability of the equations when ($z = 0$) center-to-center distance between two deuterons becomes $r_B/4 \approx 7.3 \times 10^{-13} cm$. This is only slightly more than the double deuteron radius $3.9 \times 10^{-13} cm$ where fusion occurs.

The flux, related to $\psi_1$, gets away from the surface of the caustic $A$ and moves along the $z$ axis. The flux pierces the caustic $C$ in Fig. 8(b), enters the absolute classical shadow, and reaches the Coulomb center. The flux $\psi_1$ cannot be stopped by any caustic on a way to the center since a caustic is a classical object but that flux is not bound to classical trajectories. In other words, the cusp driven state $\psi_1$ does not have classical analogues. This state is also associated with optical cusp caustics of the type shown in Fig. 6(b).

(c) Mathematical remarks

The solution $\psi_1$ of the Schrödinger equation, to the left from the caustic $A$, decays away from the $z$ axis and has the singularity $\ln(x^2+y^2)$ close to this axis. The singularity has to be compensated by the second branch of $\psi_1$ which exponentially increases from the $z$ axis and, at the first sight, destroys the scheme. But the point is that in the Coulomb field variables are separated in parabolic coordinates and the solution is a product of two wave functions which depend on an arbitrary parameter $\beta$ indicating a way of separation [11]. Therefore a wave function is a superposition of such products with different $\beta$. At a not small distance from the $z$ axis phases are large which allows to apply a saddle point integration. The path of $\beta$ integration in the second branch of $\psi_1$ is entirely on the same descent side of the saddle. In this case the integration is not reduced to the steepest descent from the saddle point and leads to a strong reduction of the result. This does not happen close to the $z$ axis where phases are small, the both branches are of the same order, and the logarithm is compensated. That is, not too close to the $z$ axis, exponentially increasing (from the $z$ axis) contributions to $\psi_1$ with different $\beta$ cancel each other due to mutual interference. In contrast, the decreasing branch of $\psi_1$ in Fig. 8(b) is not disappeared since the related $\beta$ integration is reduced to the steepest descent from the saddle point. So interference is an essential element of the phenomenon.

To the left from the caustic $A$ and not too close the $z$ axis the branch $\psi_1$ is exponentially small but $\psi_2$ is not. In principle, the former may disappear on some line directed from the cusp point in the $\{x,z\}$ - plane due to the Stokes phenomenon [17,18]. But this cannot occur since momenta along the $z$ direction, corresponding to the two branches, are finite and different which does not match conditions of Stokes lines [17,18]. So the state $\psi_1$ survives to the left from the caustic $A$.

## 4. CUSP DRIVEN TUNNELING

In classical physics the Coulomb center cannot be reached at a low energy. In quantum mechanics this depends on a form of an incident particle flux. When the flux is a usual plane wave the probability of barrier penetration is exponentially small and is generic with conventional WKB. The situation becomes different when the incident flux is of a convergent shape. In this case a cusp caustic is formed. The incident flux, restricted by the surface of this caustic (the right part of Fig. 9(a)), is accompanied by a partial leakage through the surface, marked by dashed arrows in Fig. 9(a), which corresponds to $\psi_2$.

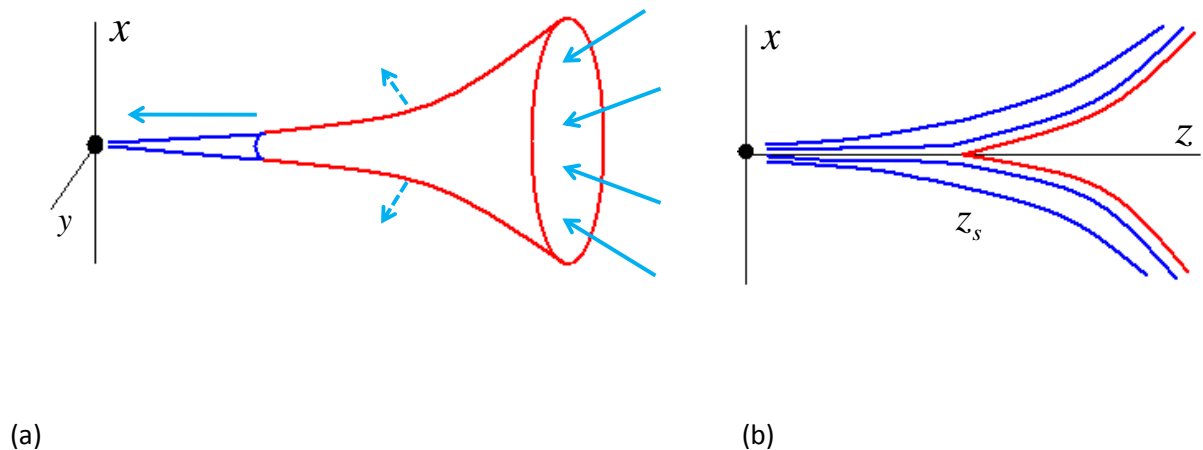

(a)                                                                         (b)

**Figure 9.** (a) The incident flux (the right part) gives rise to one directed along the channel (the left part) toward the Coulomb center. (b) In the vicinity of the cusp the flux, related to $\psi_1$, gets away from the caustic in the classically forbidden region and moves along the channel.

Unlike classical physics, there is also the flux, related to $\psi_1$, along the caustic surface which decays inside the classically forbidden region but it is not small on the surface. This flux gets away from the cusp and moves along the channel, surrounding the $z$ axis, as shown in Fig. 9(b). In this manner the flux from the cusp reaches the Coulomb center. Close to the cusp the channel width is of the order of the de Broglie wave length. Close to the Coulomb center the channel gets contracted down to the Bohr radius as shown in Fig. 9. Those estimates follow from equations (2) and (3).

The wave function inside the channel and at the caustic region is of the same value with the exponential accuracy. A reduction of the wave function at the Coulomb center, compared to the cusp point, is due to non-semiclassical effects described by preexponential factors omitted in equations (2) and (3). As a result, the probability of barrier penetration becomes not exponentially small (as $10^{-2682}$) but "normally" small. This can be called cusp driven tunneling.

A similar channel is formed by a cusp caustic in optics and acoustics where getting away from the cusp also occurs. Two branches, analogous to $\psi_2$ and $\psi_1$, interfere to the left from the cusp [19]. At large distances from the cusp, the channel is smeared out in space due to non-semiclassical effects. In our case such distances are not involved since the center is close to the cusp. Not far to the left from the cusp in Fig. 9 situations in optics and quantum mechanics are generic. But at larger distances, in contrast to optics, the rays flux $\psi_2$ is reflected by the caustic $C$. This caustic is not an obstacle for the flux $\psi_1$ since it is not connected to a set of classical trajectories. This is the reason why the flux $\psi_1$ pierces the caustic $C$ and reaches the Coulomb center.

Mechanisms, described in this paper, constitute a phenomenon of low-energy nuclear fusion. Nuclei can be of room temperature energy. Below we briefly mention experimental schemes for formation of a particle flux resulting in the cusp phenomenon.

One of experimental ways to produce the conical flux of deuterons is to confine them in a tube, for example, in a nanotube. A set of nanotubes may be explored for a wide flux of nuclei. Another way is to push deuterons (atoms) to pass through a diffraction grid of a conical shape. Since the de Broglie wave length is of the order of $1A°$, one can use a natural crystal lattice. A setup with slits can be also suitable. This is a situation of quantum lens.

From practical standpoint it is more convenient not to deal with bare deuterons but with a substance (heavy water, for example) containing them. Inside a single molecule of heavy water deuterons are in a well with vibration energy levels. An external laser radiation can influence a quantum state of deuterons in the well. One can put a question whether the radiation is able to create something like cusp state in the well and which pulse shape should be used for this purpose.